\documentclass[11pt]{amsart}

      \title[Persistent Homology Application to Hockey Analytics]
            {An Application of Topological Data Analysis to Hockey Analytics}
     \author[Daniel Goldfarb]{Daniel Goldfarb}
    \address{Niskayuna High School, 1626 Balltown Road, Niskayuna, NY 12309}
      \email{dgoldfarb15@gmail.com}
       \date{\today}

\usepackage{mathpazo}
\usepackage{amsmath,amsthm,amssymb,amsxtra,amscd,enumerate,mathrsfs,graphicx,pifont}

\usepackage{float}

\usepackage{multirow}

\usepackage{microtype}

\theoremstyle{plain}

\theoremstyle{definition}           

{\leqno{\normalshape ({\thedummy})} $$}
\numberwithin{equation}{dummy}

\theoremstyle{plain}

\theoremstyle{plain}
\newtheorem{Thm}{Theorem}[section]

\newtheorem{Cor}[Thm]{Corollary}
\newtheorem{Lem}[Thm]{Lemma}
\newtheorem{Prop}[Thm]{Proposition}

\theoremstyle{definition}
\newtheorem{Def}[Thm]{Definition}

\newtheorem{Ex}[Thm]{Example}
\newtheorem{Rem}[Thm]{Remark}

\theoremstyle{remark}
\newtheorem{Not}[Thm]{Notation}

\newcommand{\pull}
{\!\!\! -\!\!\! -\!\!\! -\!\!\!}
\DeclareMathOperator*{\hocolimprep}{hocolim}
\newcommand{\hocolim}[1]%
{\hocolimprep_{\substack{- \pull \rightarrow \\ #1}} \, }
\DeclareMathOperator*{\colimprep}{colim}
\newcommand{\colim}[1]%
{\colimprep_{\substack{- \pull \rightarrow \\ #1}} \, }

\newif\ifShowLabels
\ShowLabelstrue
\newcommand{\TeXref}[1]{
\marginpar{\scriptsize \texttt{#1}}}

\newtheoremstyle{freestylethm}{6pt}{6pt}{\itshape}{}%
                {\bfseries}{}{.5em}{\thmnote{#3}}
\theoremstyle{freestylethm}

\newcommand{\SecRef}[2]{\section{#1}\label{S:#2}%
\ifShowLabels \TeXref{{S:#2}} \fi}
\newcommand{\SSecRef}[2]{\subsection{#1}\label{SS:#2}%
\ifShowLabels \TeXref{{SS:#2}} \fi}
\newcommand{\SSSecRef}[2]{\subsubsection{#1}\label{SSS:#2}%
\ifShowLabels \TeXref{{SSS:#2}} \fi}

\newcommand{\refSS}[1]{\textup{\ref{SS:#1}}}

\newcommand{\refD}[1]{\textup{\ref{D:#1}}}

{ \begin{Thm} \label{T:#1}
\ifShowLabels \TeXref{T:#1} \fi }%
{ \end{Thm} }
\newenvironment{DefRef}[1]%
{ \begin{Def} \label{D:#1}
\ifShowLabels \TeXref{D:#1} \fi }%
{ \end{Def} }
{ \begin{Lem} \label{L:#1}
\ifShowLabels \TeXref{L:#1} \fi }%
{ \end{Lem} }
{ \begin{Cor} \label{C:#1}
\ifShowLabels \TeXref{C:#1} \fi }%
{ \end{Cor} }
{ \begin{Rem} \label{R:#1}
\ifShowLabels \TeXref{R:#1} \fi }%
{ \end{Rem} }
{ \begin{Prop} \label{P:#1}
\ifShowLabels \TeXref{P:#1} \fi }%
{ \end{Prop} }
{ \begin{Ex} \label{E:#1}
\ifShowLabels \TeXref{E:#1} \fi }%
{ \end{Ex} }
{ \begin{Not} \label{N:#1}
\ifShowLabels \TeXref{N:#1} \fi }%
{ \end{Not} }

{ \begin{Thm} [#2]\label{T:#1}
\ifShowLabels \TeXref{T:#1} \fi }%
{ \end{Thm} }
\newenvironment{DefRefName}[2]%
{ \begin{Def} [#2]\label{D:#1}
\ifShowLabels \TeXref{D:#1} \fi }%
{ \end{Def} }
{ \begin{Lem} [#2]\label{L:#1}
\ifShowLabels \TeXref{L:#1} \fi }%
{ \end{Lem} }
{ \begin{Cor} [#2]\label{C:#1}
\ifShowLabels \TeXref{C:#1} \fi }%
{ \end{Cor} }
{ \begin{Rem} [#2]\label{R:#1}
\ifShowLabels \TeXref{R:#1} \fi }%
{ \end{Rem} }
{ \begin{Prop} [#2]\label{P:#1}
\ifShowLabels \TeXref{P:#1} \fi }%
{ \end{Prop} }
{ \begin{Ex} [#2]\label{E:#1}
\ifShowLabels \TeXref{E:#1} \fi  }%
{ \end{Ex} }

\newcommand{\go}{\omega}
\newcommand{\gd}{\delta}

\DeclareMathOperator{\tun}{tun}
\DeclareMathOperator{\spar}{spar}

\ShowLabelsfalse

\begin{document}

\begin{abstract}
This paper applies the major computational tool from Topological Data Analysis (TDA), persistent homology, to discover patterns in the data related to professional sports teams.
I will use official game data from the North-American National Hockey League (NHL) 2013-2014 season to discover the correlation between the composition of NHL teams with the currently preferred offensive performance markers.   
Specifically, I develop and use the program TeamPlex (based on the JavaPlex software library) to generate the persistence bar-codes. 
TeamPlex is applied to players as data points in a multidimensional (up to 12-D) data space where each coordinate corresponds to a selected performance marker.

The conclusion is that team's offensive performance (measured by the popular characteristic used in NHL called the Corsi number) correlates with two bar-code characteristics:
greater \textit{sparsity} reflected in the longer bars in dimension 0 and lower \textit{tunneling} reflected in the low number/length of the 1-dimensional classes.
The methodology can be used by team managers in identifying deficiencies in the present composition of the team and analyzing player trades and acquisitions.
We give an example of a proposed trade which should improve the Corsi number of the team.
\end{abstract}

\maketitle



The hockey world used to be old fashioned.  Managers would recruit players strictly from what they could see with the naked eye.  Now, hockey analytics is becoming an influential cog in managing many NHL teams \cite{gS:14,jD:14}.  The Toronto Maple Leafs have already hired an Assistant Manager who is a well-known proponent of data analytics \cite{mL:14}.  In the next five years, every team is predicted to have at least one "stat analysis guru" working with them.  Unlike baseball, where it is easy to have solid position-specific stats, hockey is a faster and more fluid game. 
Hockey positions are much more dynamic and fluid, they are better described as roles that the players play.
These are shifting roles between players, especially for forwards.
The NHL keeps track of puck possession, turnovers for and against, hits for and against, shots blocked for and against, face-offs won, and scoring opportunities.  All of that in addition to some standard stats that are easy to record such as shots on goal and, of course, goals scored, goals saved, scores of games, etc.  The Dallas Stars General Manager, Jim Nill, uses a computer program from the work of 100 college students to measure Corsi numbers, turnovers, and scoring opportunities to generate statistical data for his team \cite{cP:14}. The sheer size of the data available is impressive, as can be viewed at the hockey analytics website \texttt{extraskater.com}. Even more data is publicly available on the official NHL website \texttt{nhl.com}. 

It is very unclear how to assemble this information into good use.  Only recording player stats does not give the manager the tools to assemble an effective team.  The next step should be the development of tools to analyze this data. Topology is a proper tool for taking information about individual players and generating conclusions about the team.  This is known as a local-to-global transition. This is where the Topological Data Analysis (TDA) could be useful.  This paper seems to be the first attempt to apply the major computational tool from TDA, the persistent homology, to the data collected by the NHL.

\tableofcontents

\SecRef{Geometry of Data}{Geo}

\SSecRef{Persistent homology: an overview}{perho}

The classical homology theory in topology is a way to describe, in algebraic ways, the presence and number of holes (of some dimension) in the given geometric shape or even topological space.  The precise definition and the foundations can be found in Munkres \cite{jM:00}.  The simplest algebraic invariant in a given dimension is the Betti number which expresses the "number of holes".  The computations are the simplest with the coefficients from a finite field.  For our purposes we will always use the field with two elements.

When the shape is a discrete collection of disjoint points, there is only 0-dimensional homology.  
But the points might approximate some interesting shape.  For example, if the points are in the plane then they might be tracing out some circle which can be extrapolated by a person.  
In three dimensions, the points might be lying densely on some sphere.  In these cases, the circle has non-zero 1st dimensional homology, and the sphere has non-zero 2nd dimensional homology.  Persistent homology is an idea that allows us to recognize these homology classes from the given set of disjoint points.
The articles \cite{gC:00,gCaZ:05,sW:00} give good introductions to this subject.
I will only illustrate how this works in the plane with illustrations called bar-codes created using the program TeamPlex I developed for this purpose, which extends the JavaPlex software library from the Stanford Applied Topology research group.

In the following pictures one can see the fattening of the data points happening during the construction of the \smash{\v{C}ech} complex. The details of this construction are given in section \refSS{RelH}. The emerging multiple intersections correspond to simplices in the \smash{\v{C}ech} complex.  Just for intuitive understanding, one can imagine solid disks merging together into a connected figure.  The hole that you see in Figures 3 and 4 represents the 1-dimensional persistent class that we will soon see in the javaPlex diagram. 

 \begin{figure}[H]
      \centering
      \includegraphics[width=0.7\linewidth]{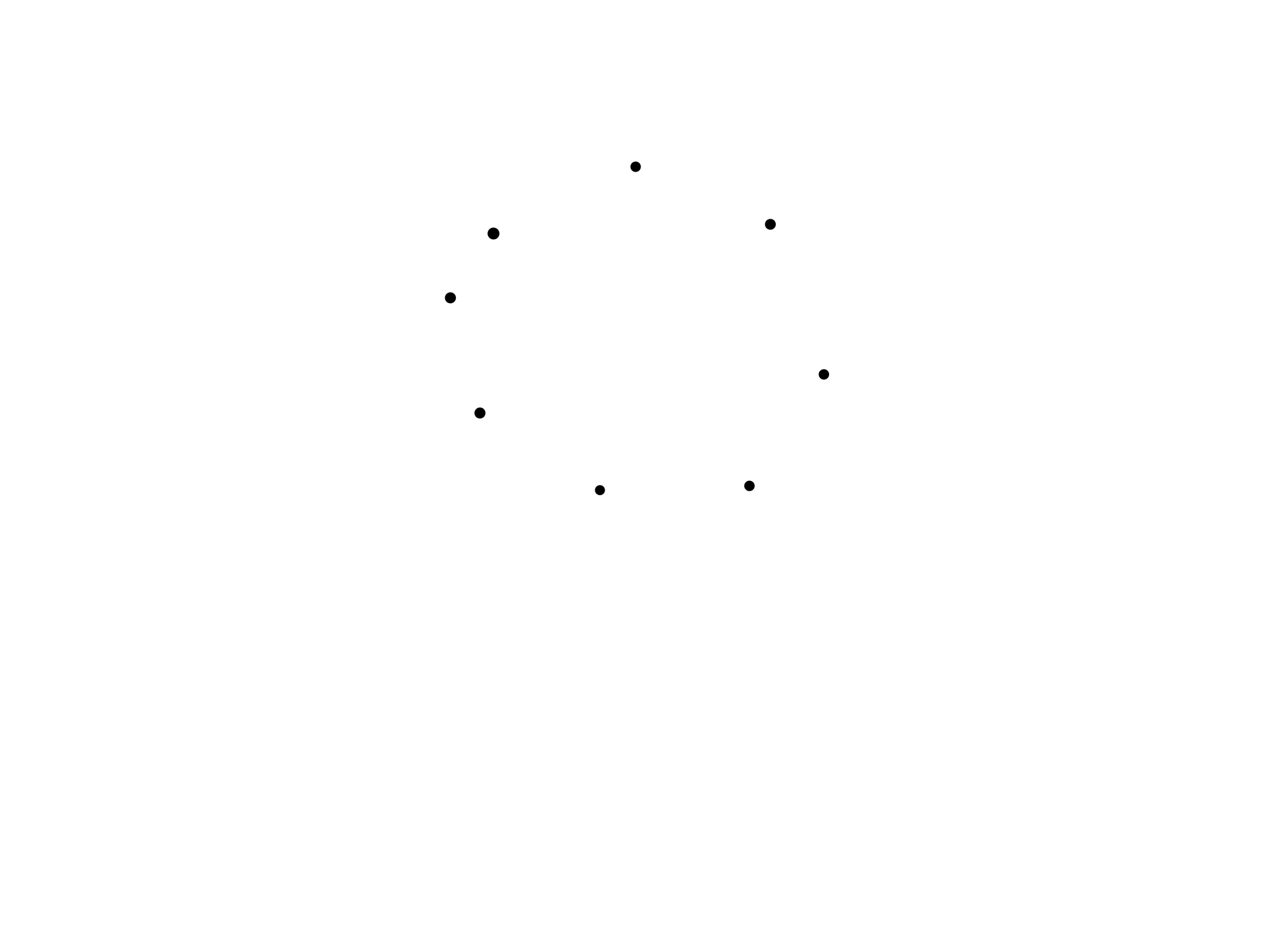}
    \caption{The discrete data set at time T=0}
      \label{photo1}
   \end{figure}
   
    \begin{figure}[H]
      \centering
      \includegraphics[width=0.7\linewidth]{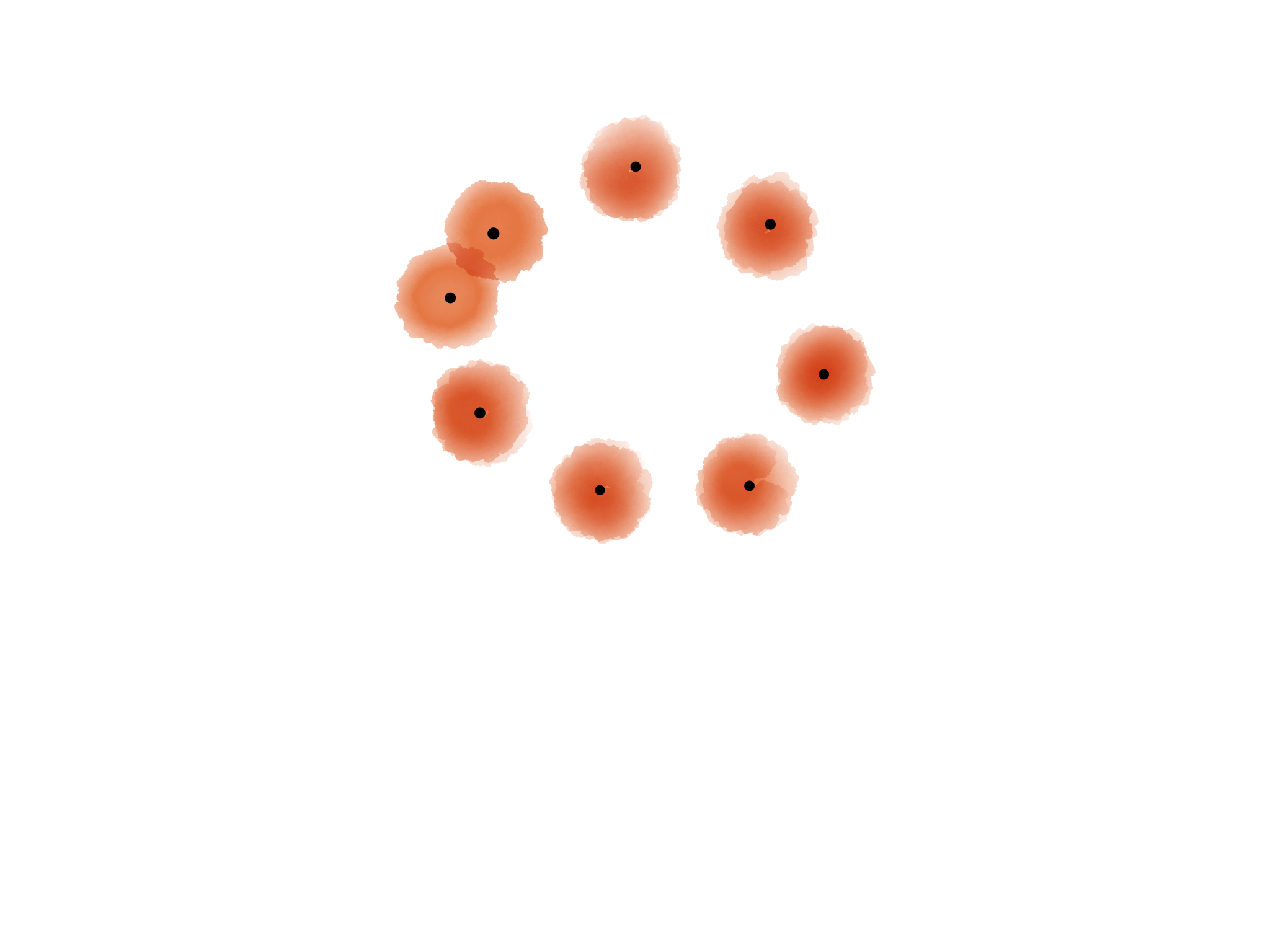}
    \caption{There is one edge formed at time T=16}
      \label{photo2}
   \end{figure}
   
    \begin{figure}[H]
      \centering
      \includegraphics[width=0.7\linewidth]{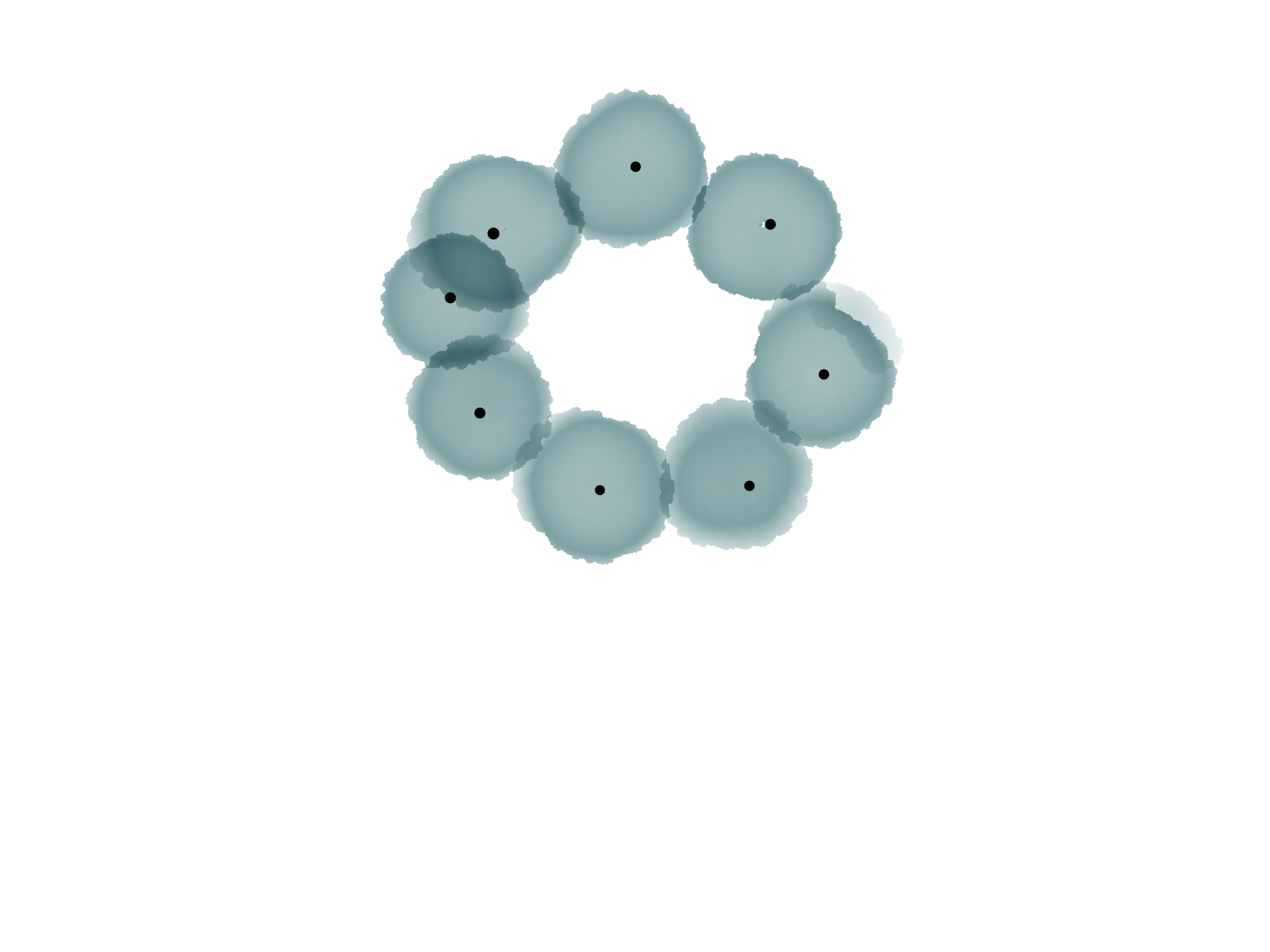}
    \caption{The first 1-dimensional cycle born at time T=60}
      \label{photo3}
   \end{figure}
   
    \begin{figure}[H]
      \centering
      \includegraphics[width=0.7\linewidth]{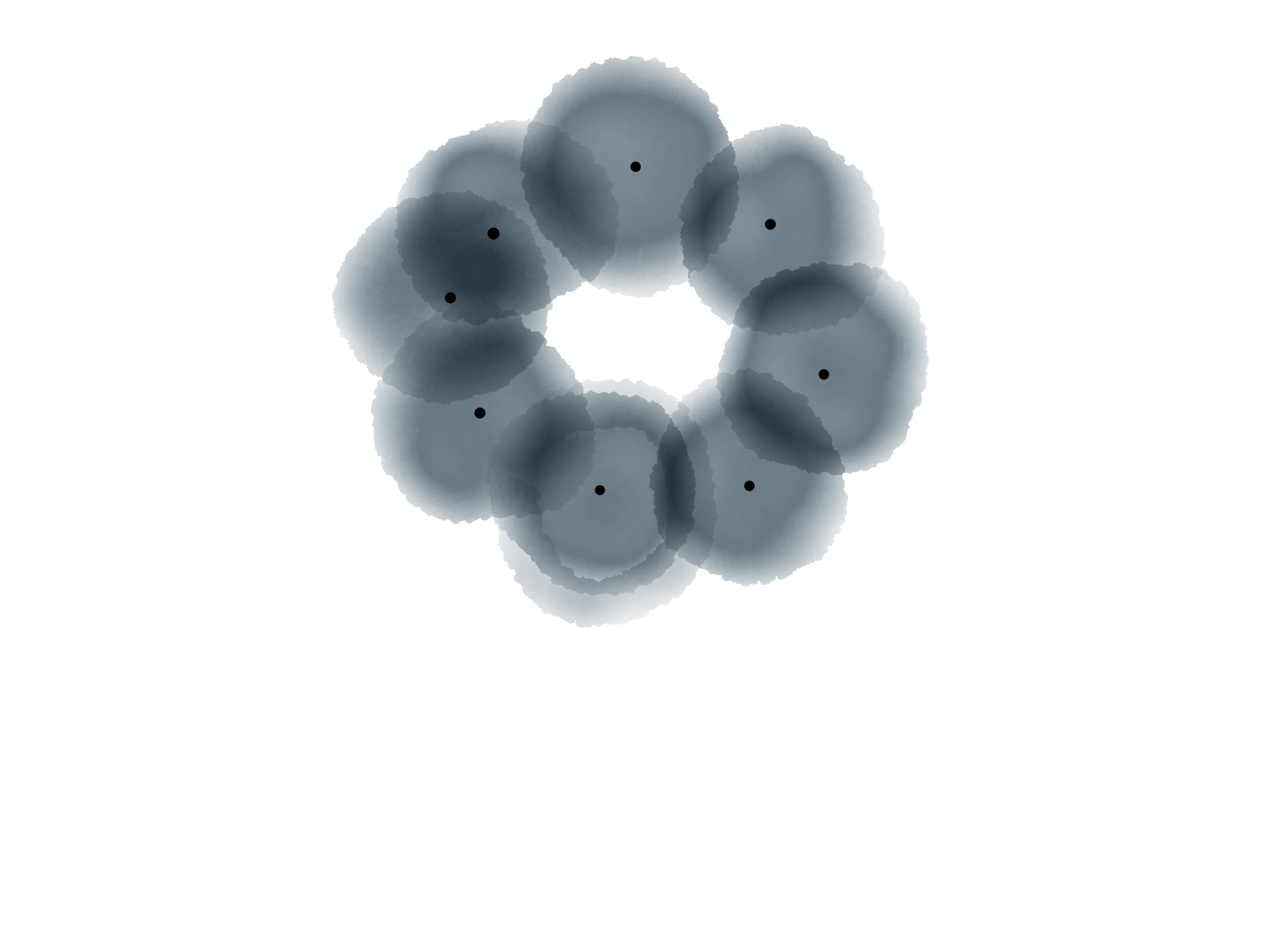}
    \caption{The 1-dimensional cycle \textit{persists} at time T=65}
      \label{photo4}
   \end{figure}
   
    \begin{figure}[H]
      \centering
      \includegraphics[width=0.7\linewidth]{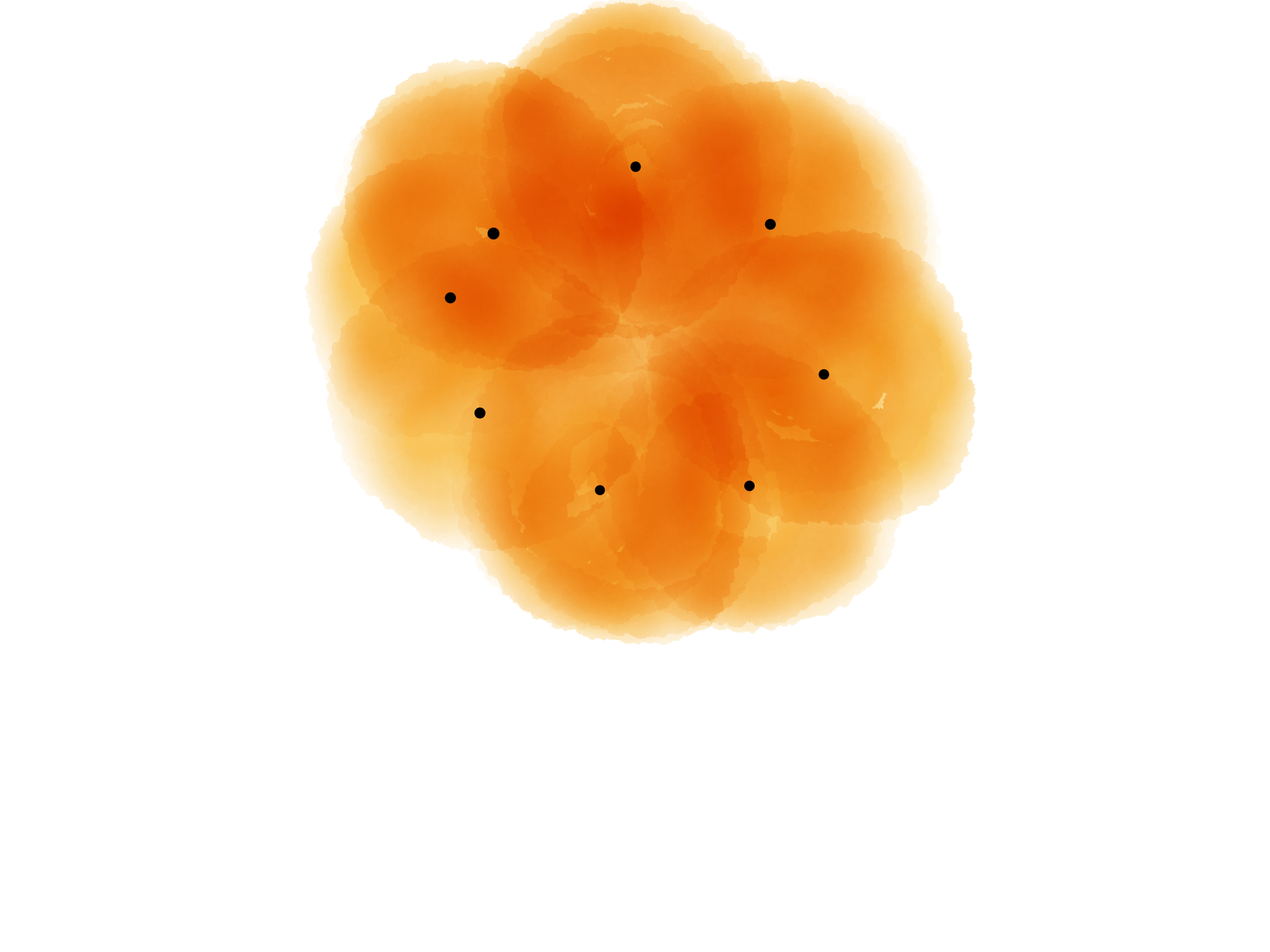}
    \caption{The complex becomes contractible at T=69}
      \label{photo5}
   \end{figure}

This whole process of fleshing out the circle from the eight points is easy to see in the 2-dimensional plane.  There are two levels of complications that appear in the application in this paper.  Even though the number of points which represent players will not increase that much, there will be at most 20 players in the data sets, the number of properties of the players will increase to 12.  This number is the number of coordinates that describe the points.  So the dimension of the space will increase from 2 to 12.  The properties of such data sets are impossible to visualize and predict.

To help with understanding the procedure, we will analyze this 2-dimen\-sional example with the tools (TeamPlex) we actually apply in 12 dimensions.

\begin{table}[H]
\centering
\begin{tabular}{lcc}
Point & X & Y\\
\hline
A & 49 & 77\\
B & 78 & 65\\
C & 90 & 32\\
D & 74 & 8\\
E & 41 & 6\\
F & 15 & 23\\
G & 9 & 48\\
H & 18 & 62\\
\end{tabular}
\caption{Eight-point data set in 2-D}
\label{2D}
\end{table}

The input data are the coordinates of the eight points in Figure 1 (named alphabetically clockwise starting with point A at the top).
TeamPlex is used to produce the following diagrams called bar-codes.

\begin{figure}[H]
      \centering
      \includegraphics[width=0.95\linewidth]{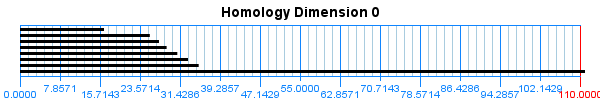}\\
      \includegraphics[width=0.95\linewidth]{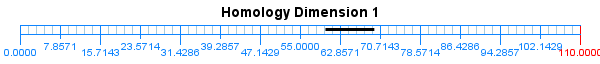} 
    \caption{0- and 1-dimensional persistent homology}
      \label{book}
   \end{figure}

This data set, as well as the team data that we use, is too small to produce any meaningful higher homology.  What we see in the first diagram is the time of the first edge being born, at the time when the first two fattened points connect.  That is the length of the first line in the diagram.  The lengths of all other lines are the times of further connections.
After time T=34, the simplicial complex is connected.

\textit{Note}: It is numerically important that there is a difference between the \smash{\v{C}ech} complex constructed in the pictures and the Rips construction used in the TeamPlex computations (see section \refSS{RelH} for details).  We ignore that distinction here, so the values of T are a little bit off.

The one single line in the second diagram is the life line of the 1-dimen\-sional class.
It roughly spans the time from Figure 3 through Figure 5.

\SSecRef{Sparsity, tunneling, and the relation to homology}{RelH}

\begin{DefRef}{thu}
Given a number $d$, take a set $S$ of points in $\mathbb{R}^d$.  This is called a \textit{point cloud}.  The \textit{cardinality} of $S$ is the number of points in $S$.  Such point cloud will be called \textit{$d$-dimensional}.
\end{DefRef}

A point cloud defines several simplicial complexes.

\begin{DefRef}{SimpComp}
Given a set $S$, a \textit{simplicial complex} with vertices $S$ is any collection $\mathcal{S}$ of subsets of $S$ that satisfies one rule: if $\Delta \in \mathcal{S}$ then for any subset $\Delta' \subset \Delta$, $\Delta' \in \mathcal{S}$.
\end{DefRef}

\begin{DefRefName}{Cech}{\v{C}ech complex}
Given a point cloud $S$ and a distance $\omega$, the \textit{\v{C}ech complex} is the simplicial complex where a subset $\Delta$ of $S$ is a simplex in $\mathcal{S}$ if there is a point $x \in \mathbb{R}^d$ such that $d(x,s) \le \omega$ for all $s \in \Delta$.  (In other words, all metric discs $D(s, \go)$, have a nonempty intersection.)
\end{DefRefName}

\begin{DefRefName}{Rips}{Rips complex}
Given a point cloud $S$ and a distance $\gd$, the \textit{Rips complex} is the simplicial complex where a subset $\Delta$ of $S$ is a simplex in $\mathcal{S}$ if for any two points in $\Delta$, say $x$ and $y$, the distance $d(x,y) \le \gd$.
\end{DefRefName}

Notice that the Rips complex construction is intrinsic for the data points and distances between them.  But the \smash{\v{C}ech} complex as in Definition \refD{Cech} is defined extrinsically because the intersections of discs are not a part of the point cloud.

Suppose we are using the \v{C}ech complexes in the persistent homology construction.
Then the radius $R$ of the largest disk that can be fit in the data without containing any points is precisely the coordinate of the death point of the 1-dimensional class.
This is only approximately true for the Rips complex.  If we only pay attention to substantially long classes, then we can ignore the difference for the simple purpose of detection.

The convex hull $[S]$ is the smallest convex set containing $S$.  In other words, $[S]$ is the intersection of all convex sets containing $S$.

\begin{DefRefName}{ZXCV}{Sparsity}
Given a point cloud $S$ in $\mathbb{R}^d$, the \textit{sparsity} of $S$ is denoted by $\spar (S)$ is the minimal distance between a pair of points contained in $S$.
The \textit{$n$-th degree sparsity} of $S$ is the $n$-vector with nondecreasing coordinates that measure progressive minimal distances between pairs of points.  So the 1st coordinate is the sparsity, the second is the shortest distance for the remaining pairs, etc.
\end{DefRefName}

This property is measured in the 0-dimensional persistent homology diagram.
From top to bottom, the lengths of lines represent the coordinates in the  $n$-th degree sparsity.

\begin{DefRefName}{Tunnn}{Tunneling}
Given a point cloud $S$ in $\mathbb{R}^d$, we will denote by $[S]$ the convex hull of $S$.  The \textit{tunneling constant} of $S$ is denoted by $\tun (S)$ and defined to be the maximal diameter of a metric ball that is contained in $[S]$ and does not intersect $S$.
\end{DefRefName}

For example, if $S$ consists of three points, then $[S]$ is a triangle and $\tun (S)$ is the diameter of the inscribed circle.  If $S$ consists of more than three points then there may not be a single inscribed circle in $S$.  Moreover, in higher dimensions, the one dimensional homology cycles that we will consider are not going to bound.  For example, suppose we have four points in $\mathbb{R}^3$.  Generically, the four points will be at the vertices of a three simplex.  A one dimensional persistent homology class can be viewed as the sum of four edges in that simplex.  In this example, the tunneling of the four points is measured by the diameter of the metric ball inscribed inside the simplex.  This tunneling correlates with the length of the persistent homology class.  The tunneling constant generalizes this intuition to higher dimensions.  The value of $\tun (S)$ guarantees that there is a point in $\mathbb{R}^d$ that contains no point from $S$ within distance $\tun (S)$.  I would like to measure this property in the team data sets.  

The reason for the term "tunneling" is that we can only detect it using the 1-dimensional homology in my application.
It is true that the most reliable computations of persistent homology are one dimensional, especially for small data sets.  In this application, the dimension $d=12$ and the cardinality of $S$ is from 14 to 20, so we can only hope to measure or estimate the property of this kind by observing 1-dimensional persistent homology.  In this case, what we measure is the width of the tunnels that exist in the data set.

In practical terms, the tunneling constant $T$ guarantees that there is a phantom point in $\mathbb{R}^d$ such that if it is added to $S$ then the tunneling constant will be cut at least in half.  

In the next section, we will generate persistence diagrams for the NHL teams and estimate the sparsity and the tunneling from the corresponding 0-dimensional and 1-dimensional components.

\SecRef{Analytics of NHL teams}{NHL}

\SSecRef{Glossary of hockey analytics terms}{Glos}  

I will need to use some terms that are common in hockey and less common terms that are used by hockey analysts.

The new trend in hockey analytics circles is that we should use primarily "shots on goal" as the measurement of team offensive quality.
The number is simple, easy to record, and reflects the fact that the team has to win possession and be aggressive in order to get close to the net in order to attempt a shot.
It is accepted that the team that has puck more often usually wins. 
This is in line with the current thinking in the NHL.

There are two slightly fancier measurements called 
Corsi and Fenwick numbers.    

Corsi or Corsi-For is the number of shot attempts by a team or player.
It is the sum of a team or players's goals, saved shots on net, shots that miss the net, and shots that are blocked. As mentioned above, it is commonly used as a proxy for puck possession.  At this point in time, no one tries to measure how long a player or team has possession of the puck, so Corsi is an approximation.  For players, the common measure is "on-ice" Corsi, or all of their team's shot attempts while they are on the ice.

Fenwick is a close relative of the Corsi number which counts unblocked shot attempts by a team or player. So it equals Corsi minus the number of shots that are blocked (by a player other than a goalie). 
Even though the consensus is that Fenwick is a prefered number, Fenwick is smaller than Corsi.
Over the limited number of games played, larger sample sizes shift the preference to Corsi for the majority of analysts.  I will use Corsi exclusively.

One more term needed is a "setup pass".
This is a pass from the player that results in the other player, who receives the pass, shooting on the net.  This is different from an assist because there is no assumption the shot is a goal.  I believe that the number of setup passes is a measurement of intent and skill rather than production.

\SSecRef{Data analysis walkthrough}{Walk}

\SSSecRef{Data collection}{DCol}

The data comes from the site \texttt{extraskater.com}.  This site contains official data from the NHL.  In order to equalize the data from all teams, we will only consider players who spend a considerable amount of time on the ice, more than 500 minutes total in the season. 
In all cases this is a number between 15 and 20.  
The statistics are taken from each team with the option "5 on 5", again to equalize the performance across the league. (This excludes the power play situations or the end-of-game pulled goalie situation when coaches use special teams.) 

For each player (row), the following statistics (column) are shown/used.

\begin{itemize}
\item[A:] name
\item[B:] goals (G)
\item[C:] assists (A)
\item[D:] setup passes (SP)
\item[E:] primary points = goals + primary assists (P1)
\item[F:] shots on goal (S)
\item[G:] corsi for (CF)
\item[H:] pass/shot ratio (PSR) taken as the percentage value
\item[I:] penalties drawn (PenDr)
\item[J:] hits for (HitF)
\item[K:] hits against (HitA)
\item[L:] take aways (Tk)
\item[M:] give aways (Gv)
\end{itemize}

The notations in parentheses are the chosen headers in the tables given on \texttt{extraskater.com}.
Let me briefly comment on the decision of which statistics to include.
The site lists lots of similar measures such as Corsi vs Fenwick numbers and lots of statistics that are too subtle such as zone start percentage.
I chose the clearly relevant statistics such as G, A. SP, S and some that reflect the personality of the player, for example PSR, PenDr, HitF, HitA.

\SSSecRef{TeamPlex application}{TPApp}

The program I wrote contains objects for the league (NHL), the teams, and the players, including a main class to run the methods in.  The statistics from extraskater.com were easily compiled into a .csv file where they were fed into the program.  
The program uses two methods from JavaPlex library.
The first method, ComputeIntervals, computes persistence classes from the coordinates of the given data cloud.
The second method, SaveBarCodesAsPNG, saves the results as a graph in png format.
I converted these files into 2-dimensional arrays in my code so they could be recognized by the JavaPlex package.  The outputs ended up in a designated folder where I was able to visually analyze the results.  I was able to implement a series of questions to fit the need of the user.  He/she can alter the number of players and stats analyzed, maximum homology dimension, and maximum filtration value.  The user can also run as many files as he/she pleases at a time.

\SSSecRef{Interpretation and conclusions}{INCO}
Across the board, the persistent homology recorded for each team consists only of 0 dimensional and 1 dimensional homology.  These are exactly the dimensions that we learned to interpret in section 1.  Recall that we expect the team to excel if its 0 dimensional homology diagram has long survival rates.  This can be seen visually in the diagram by the shaded area.  Geometrically, this represents a greater spread in qualities of players.  And we expect the team to excel if its 1 dimensional homology is short lived or non existent.  Geometrically, this represents a uniform distribution of qualities.

Observe the diagrams in the appendix.  The two best teams judged by corsi for in the 2013/2014 season were the San Jose Sharks and the Chicago Blackhawks.  We see long survival rates in the 0 dimensional persistent homology and no 1 dimensional classes.  The two worst teams were the Edmonton Oilers and the Buffalo Sabres.  The Oilers 0 dimensional diagram is the classic example of a deficient team.  The early identification at the top of the diagram represents a cluster of similar players.  Now look at the New York Islanders, the classic middle of the road team.  The 0 dimensional diagram of the islanders is precisely the average of the best and the worst teams, the sharks and the oilers.  However, the 0 dimensional diagrams of the Islanders and the Sabres (the worst team) are curiously similar.  The difference in performance is in fact reflected this time in the 1-dimensional homology.  It is non existent for the Islanders but the Sabres have two long 1 dimensional classes.  We interpret this as large tunnels in the data that make the team's composition non-uniform.  

As a practical conclusion, the team composition is satisfactory or favorable when the 0 dimensional diagram has long survival times, as in the case of the Sharks.  This seems to be the primary characteristic that can be easily visually recognized but also measured according to the length of the top line.  Another numerical characteristic of the 0-dimensional diagram is the average of the lengths of the 0-classes which is effectively the area of the shaded region.  

The secondary charactiristic is the number and length of the 1-dimensional classes.  We have seen how the solid primary qualities can be undermined by having essential "tunnels" represented by long 1-dimensional classes, as in the case of the Flyers and the Sabres.

\SSecRef{Remark}{OIU} 
The goal of this work is to predict factors that affect Corsi, a performance marker that is skewed heavily toward offense and does not reflect defensive performance (especially goaltending).  We also specialized to the "even strength" situation when the factor that is the strength of the special teams is ignored.  There might be a huge difference in the position of a team in the Corsi ranks and the formal success during the season.

To emphasize this there are two lists in Appendix C. The first column gives the Corsi rank of the team in 5 on 5 situations, the last column is the standing of the team at the end of the regular season.
The disparity between the orders of the teams in two columns is clear.

\SecRef{Discussion}{disc}

\SSecRef{Proposed explanation of the discovered correlation}{Expla}

In my application, I use what topology is best at: the transition from local to global information.
The twelve stats in the arrays are not necessarily the numerical expression of bad-to-good performance.  Simply recording player ratings will not accurately predict the success of a given team.  

The sparsity measurement is not about presence of worse and better players as much as players with diverse individual characteristics.  For example, the number of hits delivered and received and the difference between these numbers measures the style of play.  A given team is better if both the roles of (1) a skilled possession player who draws more hits and (2) the hitter who challenges possession player from the other team are represented on the team.
This seems like a natural explanation of why diversity benefits a hockey team.

Another explanation can be given from the point of view of the opponent.  
In hockey more so than in other sports, there are many one-on-one battles where it is easier to play against a team with similar properties.  You do not need to tailor your reaction or expectations.  The same goes for goalies facing shots.  It is much harder to defend against a diverse, unpredictable team.

\SSecRef{Comparison to Alagappan's NBA analytics}{comp}

I would like to compare my results to another application of TDA to sports analytics.  Two years ago, Muthu Alagappan, a Stanford engineering student, presented his research at the MIT Sloan Sports Analytics Conference \cite{jB:12}.  He analyzed the performance of NBA teams during the 2011-2012 season.  He mapped the players of the teams using the proprietary software called Mapper that was developed by the company, Ayasdi.
Alagappan \cite{mA:12} found that the more diverse a team is, the more successful it is during the season.  In his terms, successful teams had players with a variety of "positions".  In this paper, I call it sparsity.  
In basketball, and especially baseball, there are well-defined positions that make the classification and data collection easier.  Hockey positions are much more dynamic and fluid, so we cannot even talk about "many positions".  They are more like roles that the players play.
These are shifting roles between players, especially for forwards.
Paradoxically these harder to obtain statistics might be more faithful and useful in hockey.  They reflect on the kind of player you have instead of the statistics of a player asked to play a relatively static role.

The advantage of persistent homology computations compared to visually observing the Mapper diagram is that we can actually attach numerical information to observation.  For example, we can measure the life span of each homology class.

\SSecRef{Degree of success}{DSuc}
Most natural or social phenomena have a "normal" which is the most common occurrence of properties.  The created data cloud has a dense core in the middle and some flares away.  In applying persistent homology, the Rips complex quickly becomes contractible.
This means we observe no homology.  
In the usual application of persistent homology people find the normal, remove it, and focus on the peripheral properties.  

What this paper does is different, the method is applied to raw data. 
The reason that it is working so well is because the data is not natural.
The teams are made up by general managers.
Apparently we zero in on a good set of individual properties of players that persistent homology globalizes especially well.

\SSecRef{Proposed use of the analytics}{OPUA}

The general idea for application to team management is that the persistence diagrams identify deficiencies in the present composition of the team.
Once identified, the manager can attempt to eliminate the deficiencies by hiring or trade.
Ideally it would be possible to detect, for example, the center of the disk realizing the tunneling constant.  Trading for a player with known coordinates close to that center would plug that hole in the team.  This is a machine learning problem that I don't know how to solve.  Instead one could try to add available players and see if that improves the team on experimental basis.  This can be done whenever a specific trade is proposed.

Here is a hypothetical example.  Suppose we would like to improve the Buffalo Sabres team through a trade.  
After a quick experiment with possible trades, one did exceptionally well while the others made no difference in the homology.
So let us examine if a trade of Matt Ellis for Daniel Sedin of Vancouver Canucks is going to improve the team.

To make the differences evident, the following figures list the bar-codes separately in dimension 0 and dimension 1.  The first diagram is the original bar-code for the Sabres in 2013-2014, the second is the bar-code for the Sabres after the trade assuming Sedin's statistics from that season.

 \begin{figure}[H]
      \centering
      \includegraphics[width=0.95\linewidth]{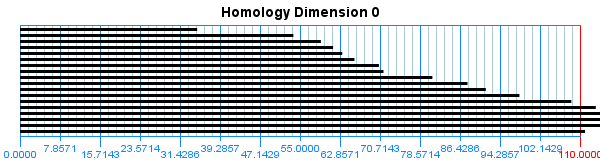}\\
      \includegraphics[width=0.95\linewidth]{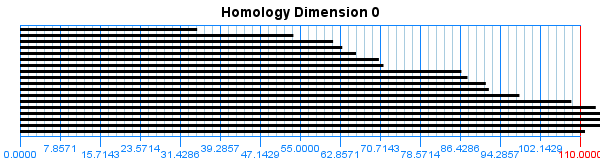} 
          \caption{Sabres vs Sabres sans Ellis plus Sedin --- dim 0}
      \label{sabdf0}
   \end{figure}

 \begin{figure}[H]
      \centering
       \includegraphics[width=0.95\linewidth]{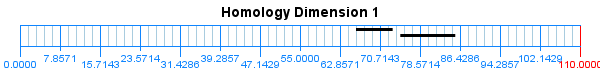}\\
      \includegraphics[width=0.95\linewidth]{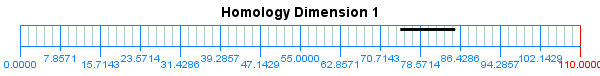} 
    \caption{Sabres vs Sabres sans Ellis plus Sedin --- dim 1}
      \label{sabdf1}
   \end{figure}

As a result of the trade, Buffalo's 1-dimensional homology lost its early cycle. One can also see the changes in the 0-dimesional bar-code that reflect a slightly better sparsity. Assuming our results are valid, this single trade between two players would result in a much more successful season for the Buffalo Sabres.

\newpage

\raggedbottom
\appendix

\section{Persistence Diagrams}

The figures show TeamFlex outcomes for several teams listed from the two best possession teams in NHL (Sharks and Blackhawks) to middle of the road teams (Canucks, Islanders, and Flyers) to the worst two teams (Oilers and Sabres). 
If the 1-dimensional classes are absent, the diagram is omitted.

   \begin{figure}[H]
      \centering
      \includegraphics[width=0.95\linewidth]{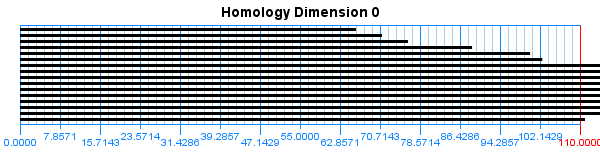}
    \caption{San Jose Sharks --- Rank 1}
     \label{sha}
   \end{figure}
   
   \begin{figure}[H]
      \centering
      \includegraphics[width=0.95\linewidth]{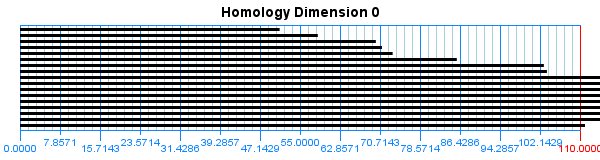}   
    \caption{Chicago Blackhawks --- Rank 2}
      \label{bla}
   \end{figure}
   
    \begin{figure}[H]
      \centering
      \includegraphics[width=0.95\linewidth]{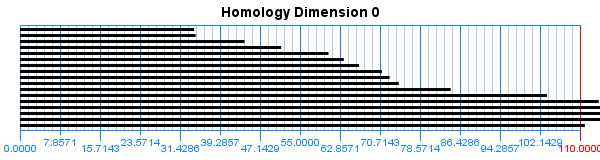}\\
      \includegraphics[width=0.95\linewidth]{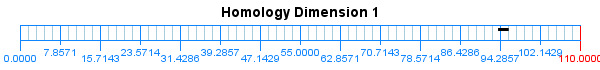} 
    \caption{Vancouver Canucks --- Rank 9}
      \label{can}
   \end{figure}
   
   \begin{figure}[H]
      \centering
      \includegraphics[width=0.95\linewidth]{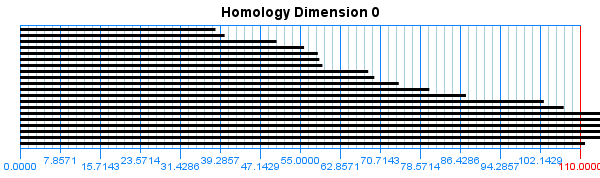}   
    \caption{New York Islanders --- Rank 11}
      \label{isl}
   \end{figure}
   
    \begin{figure}[H]
      \centering
      \includegraphics[width=0.95\linewidth]{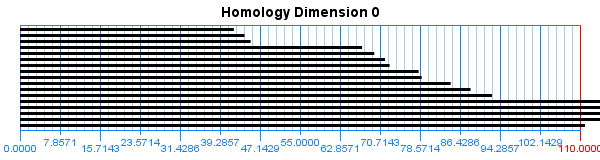}\\
      \includegraphics[width=0.95\linewidth]{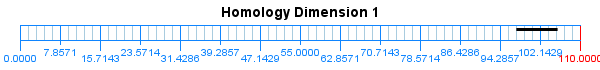} 
    \caption{Philadelphia Flyers --- Rank 15}
      \label{fly}
   \end{figure}
   
    \begin{figure}[H]
      \centering
      \includegraphics[width=0.95\linewidth]{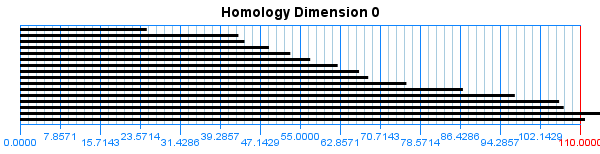}   
    \caption{Edmonton Oilers --- Rank 29}
      \label{oil}
   \end{figure}
   
   \begin{figure}[H]
      \centering
      \includegraphics[width=0.95\linewidth]{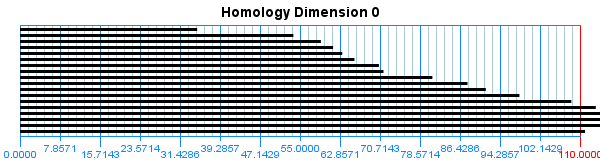}\\
      \includegraphics[width=0.95\linewidth]{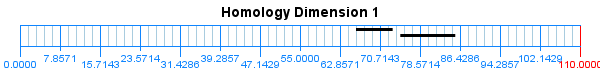} 
    \caption{Buffalo Sabres --- Rank 30}
      \label{sab}
   \end{figure}
   
   \newpage
   
   \section{Sample data files}
 
 \begin{table}[H]
\begin{tabular}{lllllllllllll}
Name & A & B & C & D & E & F & G & H & I & J & K & L \\
\hline
Justin Braun        & 3  & 13 & 114 & 8  & 99  & 227 & 0.01 & 7  & 69  & 181 & 34 & 36 \\
M.-E. Vlasic & 4  & 14 & 174 & 12 & 111 & 284 & 0.01 & 8  & 43  & 111 & 12 & 34 \\
Dan Boyle           & 5  & 8  & 132 & 9  & 90  & 177 & 2    & 11 & 39  & 90  & 22 & 47 \\
Patrick Marleau     & 15 & 23 & 371 & 30 & 172 & 300 & 120  & 6  & 79  & 72  & 40 & 47 \\
Jason Demers        & 4  & 17 & 186 & 13 & 76  & 172 & 0.01 & 10 & 61  & 112 & 30 & 57 \\
Joe Thornton        & 8  & 39 & 549 & 33 & 77  & 142 & 435  & 11 & 45  & 29  & 64 & 66 \\
Joe Pavelski        & 22 & 19 & 379 & 33 & 138 & 286 & 445  & 17 & 39  & 58  & 40 & 39 \\
Tommy Wingels       & 13 & 14 & 123 & 17 & 140 & 235 & 14   & 19 & 203 & 98  & 19 & 39 \\
Brent Burns         & 18 & 15 & 147 & 25 & 189 & 355 & 50   & 12 & 135 & 39  & 37 & 27 \\
Logan Couture       & 13 & 17 & 246 & 24 & 144 & 241 & 407  & 9  & 14  & 76  & 33 & 31 \\
Matt Nieto          & 9  & 9  & 243 & 15 & 111 & 181 & 2    & 10 & 7   & 74  & 15 & 28 \\
Tyler Kennedy       & 4  & 11 & 223 & 12 & 132 & 247 & 47   & 21 & 54  & 94  & 35 & 14 \\
A. Desjardins   & 3  & 14 & 191 & 8  & 85  & 146 & 337  & 21 & 101 & 102 & 34 & 25 \\
James Sheppard      & 4  & 15 & 202 & 12 & 83  & 141 & 94   & 9  & 103 & 111 & 31 & 15 \\
Marty Havlat        & 10 & 8  & 175 & 17 & 54  & 105 & 0.01 & 8  & 9   & 33  & 13 & 29 \\
Tomas Hertl         & 10 & 10 & 123 & 15 & 74  & 129 & 13   & 5  & 37  & 47  & 18 & 14
\end{tabular}
\caption{San Jose Sharks, best Corsi-for rank, highest sparsity}
\label{tab.sh}
\end{table}

\begin{table}[H]
\begin{tabular}{lllllllllllll}
Name & A & B & C & D & E & F & G & H & I & J & K & L \\
\hline
Jeff Petry          & 4    & 8  & 97  & 8  & 74  & 186 & 52 & 7  & 150 & 106 & 20 & 74 \\
Justin Schultz      & 9    & 10 & 142 & 15 & 63  & 138 & 103 & 6  & 25  & 101 & 14 & 52 \\
Andrew Ference      & 2    & 11 & 68  & 5  & 67  & 129 & 53 & 11 & 116 & 126 & 9  & 44 \\
Nugent-Hopkins & 11   & 20 & 230 & 22 & 115 & 197 & 117 & 12 & 62  & 67  & 44 & 39 \\
Jordan Eberle       & 17   & 21 & 306 & 30 & 138 & 250 & 122 & 15 & 49  & 110 & 43 & 54 \\
David Perron        & 17   & 20 & 288 & 30 & 165 & 279 & 103 & 21 & 107 & 104 & 46 & 33 \\
Taylor Hall         & 16   & 37 & 321 & 34 & 166 & 285 & 113 & 21 & 46  & 141 & 64 & 76 \\
Sam Gagner          & 6    & 17 & 211 & 15 & 106 & 184 & 115 & 4  & 28  & 61  & 24 & 32 \\
Nick Schultz        & 0.01 & 2  & 27  & 1  & 26  & 56  & 48 & 6  & 59  & 100 & 1  & 33 \\
Boyd Gordon         & 4    & 8  & 147 & 9  & 64  & 119 & 123 & 2  & 25  & 97  & 17 & 15 \\
Anton Belov         & 1    & 4  & 20  & 2  & 35  & 93  & 22 & 5  & 86  & 51  & 5  & 40 \\
Ryan Smyth          & 5    & 11 & 172 & 10 & 99  & 156 & 110  & 8  & 54  & 67  & 14 & 28 \\
Nail Yakupov        & 7    & 11 & 172 & 14 & 87  & 161 & 107 & 19 & 64  & 53  & 16 & 24 \\
Ales Hemsky         & 7    & 8  & 183 & 13 & 81  & 135 & 135 & 7  & 25  & 70  & 28 & 24 \\
Martin Marincin     & 0.01 & 4  & 50  & 2  & 25  & 69  & 73 & 1  & 15  & 23  & 10 & 27 \\
Mark Arcobello      & 3    & 12 & 165 & 10 & 58  & 113 & 146 & 8  & 74  & 36  & 15 & 11
\end{tabular}
\caption{Edmonton Oilers, 2nd worst Corsi-for, worst sparsity}
\label{tab.oi}
\end{table}
   
   \section{Corsi-For records from 2013-2014}

\begin{table}[H]
\centering
\begin{tabular}{lcccc}
Team & Rank & Corsi-For & Points & Standing\\
\hline
San Jose Sharks & 1 & 4089 & 111 & 5\\
Chicago Blackhawks & 2 & 3920 & 107 & 7\\
Ottawa Senators & 3 & 3892 & 88 & 21\\
Los Angeles Kings & 4 & 3888 & 100 & 10\\
Boston Bruins & 5 & 3878 & 117 & 1\\
Carolina Hurricanes & 6 & 3817 & 83 & 24\\
New York Rangers & 7 & 3815 & 96 & 12\\
Dallas Stars & 8 & 3703 & 91 & 16\\
Vancouver Canucks & 9 & 3698 & 83 & 25\\
Winnipeg Jets & 10 & 3647 & 84 & 22\\
New York Islanders & 11 & 3595 & 79 & 26\\
Phoenix Coyotes & 12 & 3585 & 89 & 18\\
Tampa Bay Lightning & 13 & 3553 & 101 & 8\\
St Louis Blues & 14 & 3547 & 111 & 4\\
Philadelphia Flyers & 15 & 3508 & 94 & 13\\
Anaheim Ducks & 16 & 3499 & 116 & 2\\
Florida Panthers & 17 & 3497 & 66 & 29\\
Nashville Predators & 18 & 3409 & 88 & 19\\
Colorado Avalanche & 19 & 3407 & 112 & 3\\
Montreal Canadiens & 20 & 3368 & 100 & 9\\
Washington Capitals & 21 & 3366 & 90 & 17\\
Columbus Bluejackets & 22 & 3348 & 93 & 14\\
Pittsburgh Penguins & 23 & 3332 & 109 & 6\\
Calgary Flames & 24 & 3321 & 77 & 27\\
Detroit Red Wings & 25 & 3309 & 93 & 15\\
Toronto Maple Leafs & 26 & 3259 & 84 & 23\\
New Jersey Devils & 27 & 3223 & 88 & 20\\
Minnesota Wild & 28 & 3154 & 98 & 11\\
Edmonton Oilers & 29 & 3103 & 67 & 28\\
Buffalo Sabres & 30 & 3009 & 52 & 30\\
\hline
\end{tabular}
\caption{NHL season 2013-2014}
\label{tab}
\end{table}

\newpage

\raggedbottom

\end{document}